\documentclass[a4paper,12pt]{article}

\usepackage{amsmath}
\usepackage[pdftex]{graphicx}
\usepackage{multirow}
\usepackage{gensymb}
\usepackage{float}
\usepackage{graphicx}
\usepackage[hidelinks]{hyperref}
\usepackage{caption}
\usepackage[table,xcdraw]{xcolor}
\usepackage[left=2cm,right=2cm,top=2cm,bottom=2.5cm,head=1cm,foot=1cm]{geometry}
\usepackage[utf8]{inputenc}
\usepackage[magyar,english]{babel}
\usepackage[backend=biber,sorting=none,doi=false,url=false,style=numeric-comp]{biblatex}
\newcommand{\sqsntwo}{$\sqrt{s_{_{\text{NN}}}}=2.76$~TeV }

\newcommand{\mT}{m_{\text{T}}}

\newcommand{\kT}{k_{\text{T}}}

\author{Balázs Kórodi$^1$, Dániel Kincses$^1$, Máté Csanád$^1$,\\
            {\small $^1$Eötvös Loránd University, Budapest, Hungary}}
                  
\title{Event-by-event investigation of the two-particle source function in $\sqrt{s_{_{\text{NN}}}}=2.76$~TeV PbPb collisions with EPOS}
\date{\today}

\begin{document}

\begin{titlepage}
\maketitle
\begin{abstract}
The investigation of the two-particle source function in lead-lead collisions simulated by the EPOS model at a center of mass energy per nucleon pair of $\sqrt{s_{_{\text{NN}}}}=2.76$~TeV is presented. The two-particle source functions are reconstructed directly on an event-by-event basis for pions, kaons and protons separately, using the final stage of EPOS. A Lévy source shape is observed for all three particle species in the individual events, deviating significantly from a Gaussian shape. The source parameters are extracted as functions of collision centrality and pair average transverse mass ($m_{\text{T}}$). The Lévy exponent is found to be ordered accordingly to particle mass. The Lévy scale parameter is found to scale for all particle species with $m_{\text{T}}$ according to Gaussian hydrodynamic predictions; however, there is no $m_{\text{T}}$-scaling found across these species. In case of pions, the effects of the decay products and hadronic rescattering are also investigated. The Lévy exponent is decreased when decay products are also included in the analysis. Without hadronic rescattering and decay products, the source shape is close to a Gaussian.
\end{abstract}
\end{titlepage}

\section{Introduction}

Since the discovery of the quark-gluon plasma~(QGP), an important goal of high-energy heavy ion physics has been to understand the space-time geometry of the medium created in the collisions~\cite{Lednicky:2001qv,Lisa_2005}. The main quantity under investigation is the space-time probability density $S(x,p)$ of the particles after the kinetic freeze-out, where $x$ and $p$ are the position and momentum four-vectors of a given particle at its creation, respectively. The $S(x,p)$ distribution is often referred to as the single-particle source function. The precise form of this function is important for a multitude of reasons. It is connected to the hydrodynamic expansion of the medium~\cite{makhlin1988hydrodynamics,Csorg__1996}, it appears in studies of light nuclei formation~\cite{Oliinychenko:2020ply}, and it can even be a sign of critical phenomena~\cite{Csorgo:2005it}.

The experimental investigation of $S(x,p)$ is possible, although not with direct methods, since the length scale of the process is on the femtometer level. The sub-field of high-energy physics dedicated to the measurement of femtometer scale structures is called femtoscopy. The field originates back to the 1950s, when it was shown that information about the source function can be obtained by measuring momentum correlations of pions~\cite{goldhaber1959pion,PhysRev.120.300}. The effect that is used is analogous to the HBT-effect~\cite{HanburyBrown:1952na} known from astronomy, caused by the quantum-statistical nature of bosons and fermions. Over the years, many experimental analyses have been performed, and today femtoscopy is still one of the extensively studied subjects of high-energy heavy ion physics. 

In early measurements, a Gaussian source shape was assumed, providing an acceptable description of the data~\cite{Adler_2004,STAR:2004qya}. These results coincided with the prediction of hydrodynamic model calculations~\cite{Csorg__1996,Akkelin:1995gh,Csorgo:1994fg}, where the Gaussian source is obtained as the result of the hydrodynamic expansion of the QGP medium. On the other hand, source imaging studies~\cite{PHENIX:2007grx,PHENIX:2006nml} suggested that the source has a long-range component, obeying a power-law behavior. Recent high precision correlation measurements~\cite{Adare_2018} at RHIC have shown that the pion source in central gold-gold collisions at a center of mass energy per nucleon pair of $\sqrt{s_{_{\text{NN}}}}=200$~GeV does not follow a Gaussian distribution, but instead a so-called L\'evy alpha-stable distribution is needed to adequately describe the data. In recent years, it was found that the source has a L\'evy shape for various centralities, collision energies, and colliding systems~\cite{Kincses:2017zlb,Lokos:2018dqq,Porfy:2023yii,NA61SHINE:2023qzr,CMS:2023xyd} as well.

There are many potential reasons behind the appearance of L\'evy distributions in high-energy heavy ion collisions. Anomalous diffusion in the expanding medium after hadronization would lead to a L\'evy distributed source~\cite{Csorgo:2003uv, Csanad:2007fr} due to the generalization of the central limit theorem. Further plausible physical reasons are jet fragmentation~\cite{Csorgo:2004sr}, resonance decays~\cite{Kincses:2022eqq} and the vicinity of the critical point~\cite{Csorgo:2005it}. It was also shown recently~\cite{Cimerman:2019hva} that event averaging and non-sphericity could also lead to the appearance of a non-Gaussian component in experimental measurements where many events are averaged. To understand the underlying processes leading to the appearance of the L\'evy distribution, additional efforts are needed from the phenomenology side. An available option is to use event generators, such as EPOS~\cite{Werner:2010aa} to simulate high-energy heavy ion collisions. In EPOS, the source function can be directly reconstructed, therefore various investigations can be performed that are not accessible using experimental data. 

In this paper, a detailed event-by-event investigation of the pion, kaon and proton two-particle source function is presented using lead-lead (PbPb) collision data at $\sqrt{s_{NN}} = 2.76$ TeV simulated by the EPOS model. The reconstructed two-particle source functions are fitted with a three-dimensional, spherically symmetric L\'evy distribution. The source parameters are studied as functions of pair transverse mass, particle type and collision centrality.

The paper is structured as follows. In Sec.~\ref{s:twoparicletsource} the basic properties of the two-particle source function and those of L\'evy distributions are discussed. Subsequently, in Sec.~\ref{s:analysis} the details of our analysis and methods are given. The results on the Lévy parameters as a function of pair transverse mass, collision centrality, and particle type are discussed in Sec.~\ref{s:results}. Finally, a summary is presented in Sec.~\ref{s:summary}.

\section{The two-particle source function} \label{s:twoparicletsource}

The two-particle (pair) source function $D(r,K)$ is defined as the auto-correlation of the single-particle source $S(x,p)$ in the spatial variable at identical single-particle momenta:
\begin{equation} \label{e:Dr}
    D(r,K) = \int S(x_1,K) S(x_2,K)d^4\rho = \int S(\rho + \frac{r}{2},K)S(\rho-\frac{r}{2},K)d^4\rho,
\end{equation}
where $x_1$ and $x_2$ are the position four-vectors of the two particles, $\rho = (x_1 + x_2)/2$ is the pair center-of-mass four-vector, $r=x_1 - x_2$ is the pair separation four-vector and $K=(p_1 + p_2)/2$ is the average momentum of the pair, with $p_1$ and $p_2$ being the individual four-momenta of the two particles. The pair-source $D(r,K)$ can be reconstructed indirectly from momentum correlation measurements and is also usually directly accessible in simulations. As both $r$ and $K$ are four-vectors, $D(r,K)$ generally has 8 variables. In the usual procedure described in e.g.\ Ref.~\cite{Adare_2018}, the dependence on $r$ is parameterized using some source parameters, and the dependence on $K$ is then carried by those parameters. Based on this, the $K$-dependence is suppressed in the notation, while one keeps in mind that the parameters of $D(r)$ vary with $K$. In case of a spherically symmetric Gaussian source, $D(r)$ is then given by:
\begin{equation}
    D(r)\propto e^{-\frac{\left |\boldsymbol{r}\right |^2}{4R(K)^2}},
\end{equation}
where $R(K)$ is the momentum-dependent scale parameter, also called homogeneity length or HBT radius. The time dependence of $D(r)$ and the mass-shell condition for the momenta can be coupled and when calculating momentum-correlation functions, $D(r)$ can usually be considered to depend only on the three-vector $\boldsymbol{r}$~\cite{Adare_2018}. Furthermore, the source is often assumed to be spherically symmetric in a given reference frame which means that it only depends on $|\boldsymbol{r}|$.

\begin{figure}[h!]
\centering
\includegraphics[trim=0 0 0.5cm 0, clip,scale=0.6]{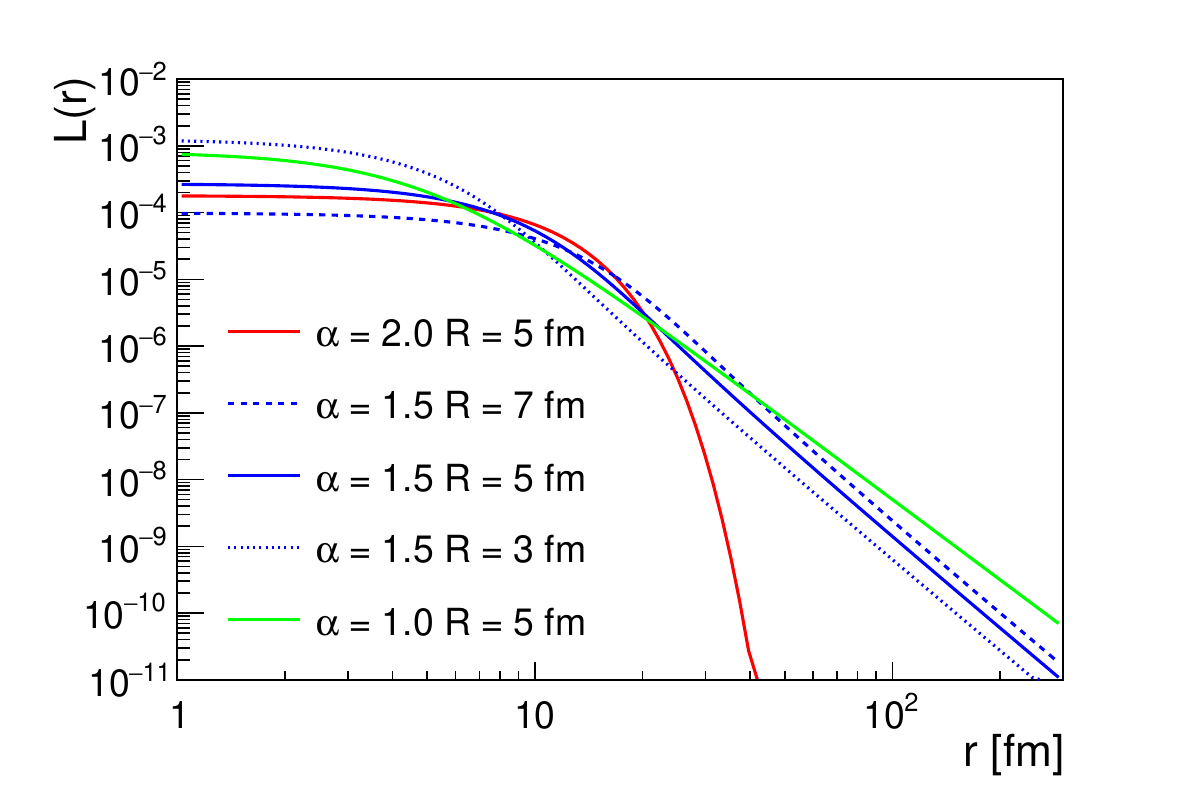}
\caption{The one-dimensional projection $\mathcal{L}_{1\rm{D}}(r;\alpha,R)$ of the three-dimensional L\'evy distribution for different parameter values.}
\label{f:Levyfunc}
\end{figure}

As mentioned above, for the description of precise momentum correlation measurements, L\'evy distributions are required instead of the special case of Gaussian distributions. The symmetric L\'evy alpha-stable distribution is a generalization of the Gaussian distribution defined by the following Fourier transform in three dimensions:
\begin{equation} \label{e:levy}
    \mathcal{L} (\boldsymbol{r};\alpha,R)=\frac{1}{(2\pi)^3}\int e^{i\boldsymbol{qr}}e^{-\frac{1}{2}|\boldsymbol{q}R|^{\alpha}}d^3\boldsymbol{q},
\end{equation}
with $\boldsymbol{q}$ being an arbitrary integration variable vector. The two parameters of the distribution are the L\'evy stability index $\alpha$ and the L\'evy scale parameter $R$. The Gaussian distribution is obtained with $\alpha=2$ and the Cauchy distribution is obtained with $\alpha=1$. The power-law tail appears for $\alpha<2$, as shown in Fig.~\ref{f:Levyfunc}. In case of $\alpha \leq 2$ the distribution is stable, meaning that it represents the limit of properly normalized sums of random variables.  If the source has a L\'evy shape with $\alpha \leq 2$, then the two-particle source also has a L\'evy shape with the same $\alpha$, but with different $R$:
\begin{equation}
    S(\boldsymbol{x})=\mathcal{L}(\boldsymbol{x};\alpha ,R)\Rightarrow D(\boldsymbol{r})=\mathcal{L}(\boldsymbol{r};\alpha ,2^{1/\alpha}R).
\end{equation}
A value of $\alpha > 2$ is not physically allowed for a stable L\'evy distribution, as shown in Ref.~\cite{UchaikinZolotarev+2011}. The $R$ parameter is simply the quadratic mean of the distribution in the Gaussian case. For $\alpha<2$ however, the L\'evy distribution does not have a second moment, but $R$ is still proportional to the full width at half maximum (with the proportionality constant depending on $\alpha$), hence it represents the spatial scale.

The one-dimensional projection of a spherically symmetric three-dimensional L\'evy distribution can be calculated by performing the integrals for the angular components of $\mathbf{q}$ in Eq.~(\ref{e:levy}) and averaging over the angles of $\mathbf{r}$, resulting in the following expression:
\begin{equation} \label{e:levy1d}
    \mathcal{L}_{1\rm{D}}(r;\alpha,R)=\frac{1}{2r\pi^2}\int_0^{\infty}q\sin(qr) e^{-\frac{1}{2}(qR)^{\alpha}}dq.
\end{equation}
In Fig.~\ref{f:Levyfunc}, $\mathcal{L}_{1\rm{D}}(r;\alpha,R)$ is shown for different parameter values. The tail of the distribution is heavier for smaller values of $\alpha$. Distributions with the same $\alpha$ share the same asymptotic behavior, i.e., the power-law exponent for large $r$. The $R$ parameter controls the width of the distribution (and, given the normalized nature of the shown distributions, it also affects the value at $r=0$).

The importance of making the correct source shape assumption lies in the extraction of the source parameters: clearly a Gaussian radius is not the same as a L\'evy radius, and if the former does not describe the data in a statistically acceptable manner then the source parameter is not a relevant descriptor of the data. Equally importantly, for a source resembling a L\'evy shape, the extracted Gaussian radii entangle the L\'evy radius and the L\'evy exponent, and the possibly non-monotonic change of both with collision energy may be hidden. The interest of measuring the correlation strength~\cite{Vance:1998wd,Csorgo:1999sj,Csorgo:2009pa,Adare_2018} also underlines the importance of making the statistically acceptable assumption for the source shape. Finally, the value of the L\'evy exponent itself could be important for several physical phenomena~\cite{Csorgo:2003uv, Csanad:2007fr, Csorgo:2004sr, Csorgo:2005it}.

\section{Analysis details}\label{s:analysis}

In our analysis, we utilized the EPOS phenomenological model~\cite{Werner:2010aa}. EPOS can be used to simulate high-energy heavy ion and proton-proton collisions. It reproduces all the basic experimentally measured quantities near zero bariochemical potential ($\mu_{\text{B}}$), corresponding to top RHIC or LHC energies. In EPOS, a merged version of the Gribov-Regge theory and the eikonalized parton model is implemented~\cite{Drescher:2000ha}. In order to account for the variable local string density in different regions of the medium, string segments (based on local density as well as transverse momenta of the segments) are divided into the \textit{core} and the \textit{corona}~\cite{Werner:2010aa, Werner:2007bf, Werner:2013tya}. Subsequently, a 3+1 dimensional viscous hydrodynamic-based algorithm is applied~\cite{Werner:2010aa}; while the equation of state X3F was utilized, retaining compatibility with lattice calculations~\cite{Borsanyi:2010cj} at $\mu_{\text{B}}=0$. At the final stage, the Cooper-Frye formula~\cite{Cooper:1974mv} is utilized, and a \textit{hadronic afterburner} is used, based on UrQMD~\cite{Bass:1998ca, Bleicher:1999xi}.

For the analysis presented below, approximately $8.0\times10^5$ PbPb events were simulated by the EPOS model at $\sqrt{s_{_{\text{NN}}}}=2.76$~TeV, using version 3.451. The data set contained the mass, the momentum, and the position four-vector for each particle at the kinetic freeze-out. Further information about the particles e.g. whether they are decay products or not, was also available.

For each individual event the centrality, describing the degree of overlap of the two incident nuclei, was calculated via an ordering of all simulated events based on the impact parameter. As usual, 0\% centrality corresponds to the most central events and 100\% to the most peripheral ones.

\subsection{Reconstruction of the two-particle source function}

Prior momentum correlation measurements~\cite{Adler_2004,STAR:2004qya,PHENIX:2007grx,PHENIX:2009ilf,STAR:2009fks,CMS:2017mdg} have shown that the two-particle source function is approximately spherically symmetric in the longitudinally co-moving system of the pair (LCMS), where the longitudinal component of the pair average momentum is zero. The LCMS pair separation vector $\boldsymbol{r}_{\text{LCMS}}$ can be expressed with the lab-frame single-particle coordinates as
\begin{equation}
    \boldsymbol{r}_{\text{LCMS}}=\left (x_1 - x_2, y_1 - y_2,  \frac{z_1 - z_2 -\beta (t_1 - t_2)}{\sqrt{1-\beta ^2}} \right ), 
\end{equation}
\begin{equation}
    \beta = \frac{p_{z,1} + p_{z,2}}{E_1 + E_2},
\end{equation}
where $x_i, y_i, z_i$ are the lab-frame coordinates at the kinetic freeze-out, $t_i$ is the time of the kinetic freeze-out, $p_{z,i}$ is the $z$ component of the momentum and $E_i$ is the energy for the two particles indexed as $i=1,2$. The distribution of two-particle distances in terms of $\boldsymbol{r}_{\text{LCMS}}$ is in fact $D(\boldsymbol{r}_{\text{LCMS}})$, where the $K$ and time dependence is suppressed in notation according to Sec.~\ref{s:twoparicletsource}. In the analysis, a spherically symmetric source is assumed, and the distribution of $r_{\text{LCMS}}\equiv \left |\boldsymbol{r}_{\text{LCMS}} \right |$, denoted by $D^{(\text{ang.av.})}(r_{\text{LCMS}})$, is measured. The superscript (ang.av.) means that this distribution is obtained by averaging $D(\boldsymbol{r}_{\text{LCMS}})$ over the full solid angle $\Omega _{\text{LCMS}}$:
\begin{equation}
    D^{(\text{ang.av.})}(r_{\text{LCMS}})=\frac{1}{4\pi}\int D(\boldsymbol{r}_{\text{LCMS}})r_{\text{LCMS}}^2 d\Omega _{\text{LCMS}}.
\end{equation}
Because of the spherical symmetry, $D(\boldsymbol{r}_{\text{LCMS}})$ does not depend on $\Omega _{\text{LCMS}}$, thus the integral results in a $4\pi$ multiplicative factor. The one-dimensional two-particle source distribution $D(r_{\text{LCMS}})$ is then obtained with:
\begin{equation}\label{e:dimtot}
    D(r_{\text{LCMS}})=\frac{D^{(\text{ang.av.})}(r_{\text{LCMS}})}{r_{\text{LCMS}}^2}.
\end{equation}

In the analysis, $D(r_{\text{LCMS}})$ was measured on an event-by-event basis for pion, kaon, and proton pairs separately. Each individual particle was required to have $|\eta|<1$, where $\eta$ is the pseudorapidity, to facilitate comparison with experimental results, as well as to focus on midrapidity particle production. In case of pions, pairs with positive-positive, negative-negative, and positive-negative electric charges were treated separately. This separation was not possible for kaon and proton pairs due to the limited number of such particles in a single event. The measurement was performed in average pair transverse momentum ${\kT=\sqrt{K_x^2 + K_y^2}}$ classes. The number and the size of the $\kT$ classes for the different particle species were determined by the average number of pion, kaon, and proton pairs in the individual events. Based on this, 10 $\kT$ classes in the range of 250--1000~MeV/$c$ were defined for pions, 5 classes in the range of 200--1000~MeV/$c$ for kaons, and 3 classes in the range of 200--1400~MeV/$c$ for protons. Furthermore, events were categorized based on their centrality: 4 centrality classes (0-5\%, 5-10\%, 10-20\%, 20-30\%) were investigated. In more peripheral collisions the number of pairs is significantly lower, therefore only the mentioned central classes were used. Altogether $8.0\times10^5$ events were analyzed, resulting in approximately $4.0\times10^4$ and $8.0\times10^4$ events in the 5\% and 10\% wide centrality classes, respectively.

As detailed in the previous section, EPOS has different stages of evolution. In this analysis, both the core and the corona, as well as the hadronic rescattering governed by UrQMD were utilized for the final results. In case of pions, four separate cases were investigated. In the first two cases, only the core part was utilized without UrQMD, while in the third and fourth cases, both the core and the corona were used together with UrQMD. The first and the third cases included only primordial pions coming directly from the medium, while the second and the fourth cases included the decay products as well. For kaons and protons only the fourth case was studied, because the number of pairs in a single event was not sufficiently high without the inclusion of the decay products and without UrQMD. 

\subsection{Fitting the two-particle source function}

\begin{figure}[h!]
\centering
\includegraphics[trim=0 0 0.5cm 0, clip,scale=0.9]{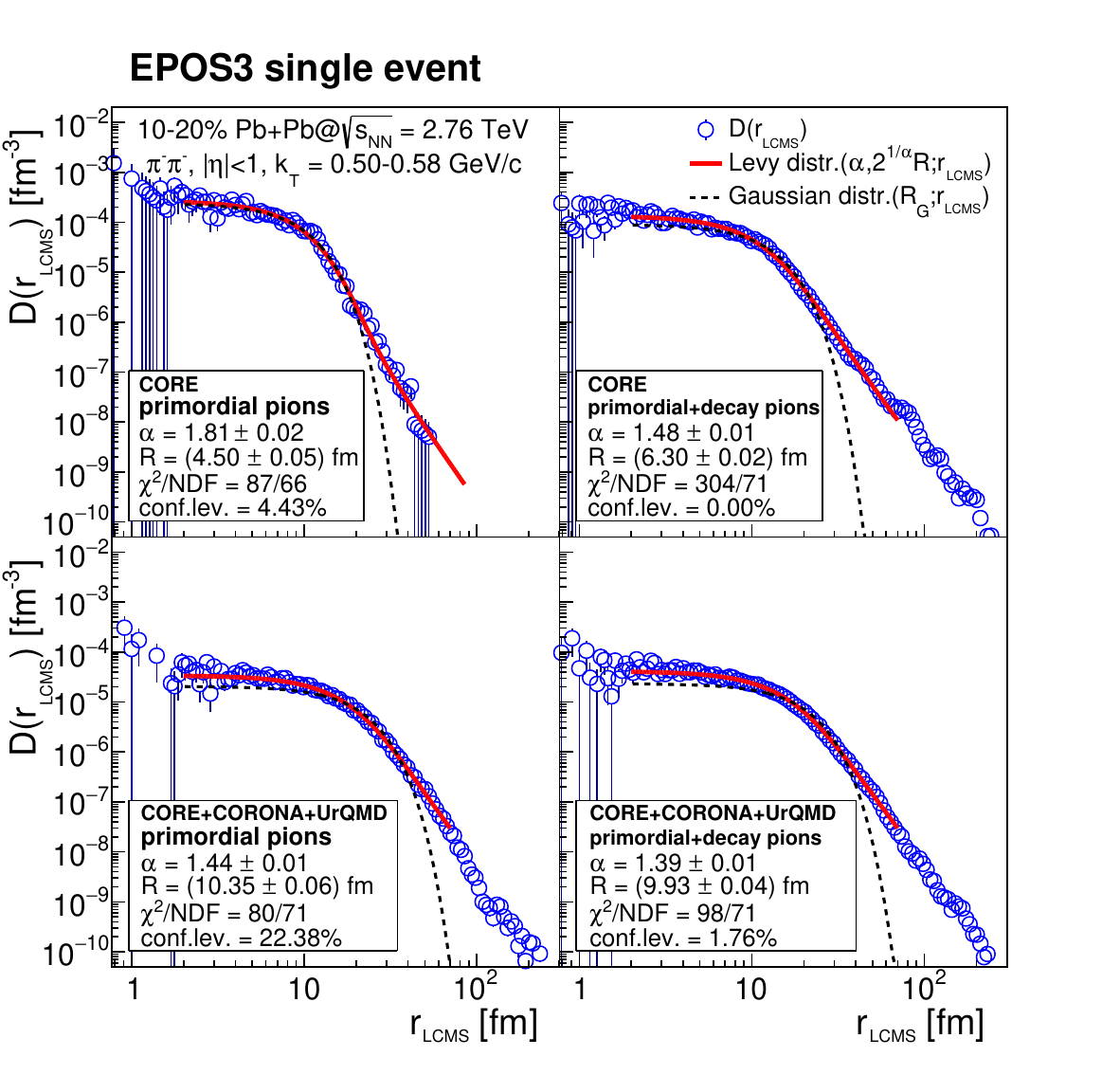}
\caption{Example L\'evy and Gaussian fits to the one-dimensional two-particle source function $D(r_{\text{LCMS}})$ reconstructed from a single event in the centrality range of 10--20\%, using negatively charged pion pairs with $\kT=0.50-0.58$~GeV/$c$. The four investigated cases are CORE primordial pions (upper left), CORE primordial+decay pions (upper right), CORE+CORONA+UrQMD primordial pions (lower left), and CORE+CORONA+UrQMD primordial+decay pions (lower right). The L\'evy fits correspond to the red curves, while the Gaussian fits correspond to the dashed black curves. The L\'evy fit parameters are shown in the individual panels.}
\label{f:pionlevyfit}
\end{figure}

\begin{figure}[h!]
\centering
\includegraphics[trim=0 0 0.5cm 0, clip,scale=0.9]{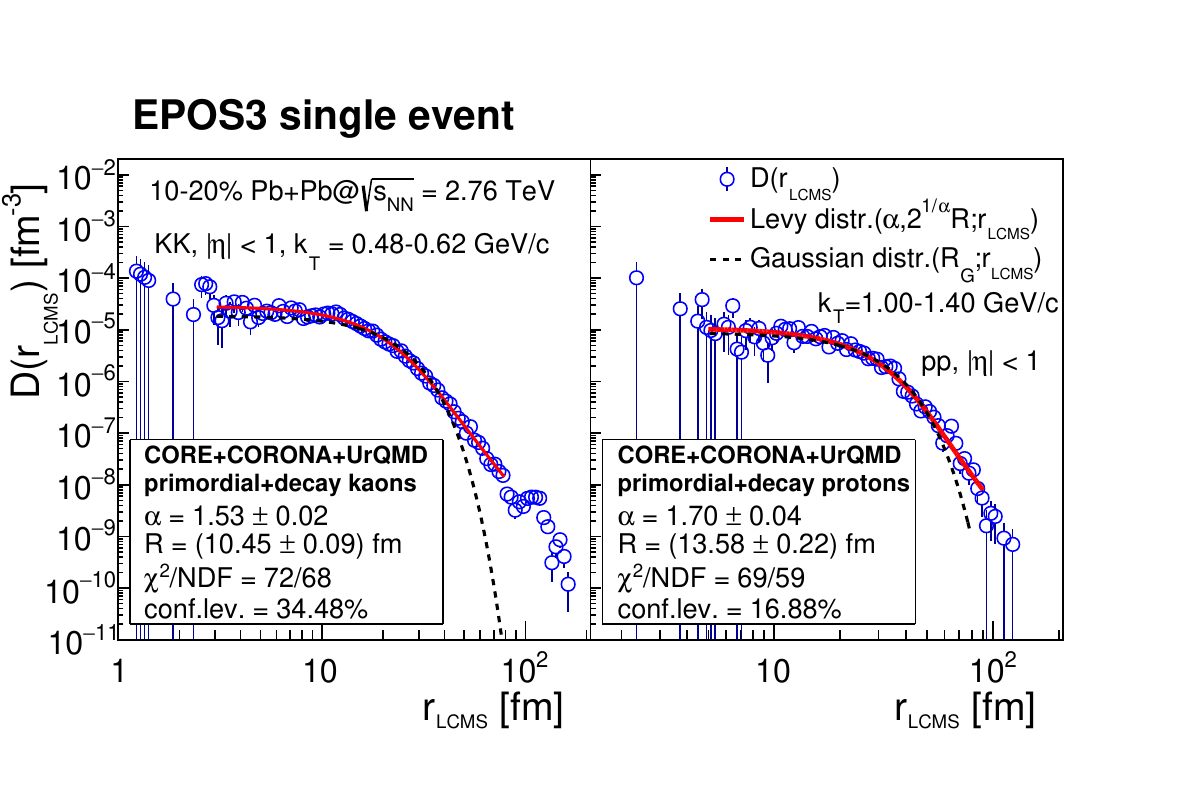}
\caption{Example L\'evy and Gaussian fits to the one-dimensional two-particle source function $D(r_{\text{LCMS}})$ reconstructed from a single event in the centrality range of 10--20\%, using CORE+CORONA+UrQMD primordial+decay kaons (left) and protons (right). The L\'evy fits correspond to the red curves, while the Gaussian fits correspond to the dashed black curves. The L\'evy fit parameters and the $\kT$ ranges are shown in the individual panels.}
\label{f:kaonprotonlevyfit}
\end{figure}

The $D(r_{\text{LCMS}})$ distributions were calculated based on the simulations via a logarithmic binning, and a normalization by the event multiplicity was applied. L\'evy and Gaussian fits were performed for each individual event, as shown in the example fits in Figs.~\ref{f:pionlevyfit} and~\ref{f:kaonprotonlevyfit}. The L\'evy fit function $\mathcal{L}_{1\rm{D}}(r;\alpha,R)$ was the one-dimensional projection of the three-dimensional L\'evy distribution, calculated in Eq.~(\ref{e:levy1d}) and shown in Fig.~\ref{f:Levyfunc} for different parameter values. A multiplicative normalization factor $N$ was also included in the fits, but due to the pre-normalization, this factor turned out to be very close to unity in all cases. The fits were performed using standard $\chi^2$ minimization. The lower fit limit was 2~fm, 3~fm, and 5~fm for pions, kaons, and protons, respectively. The upper fit limit was varied between 70~fm and 100~fm for all three particle species, and the value resulting in the highest confidence level was retained. There were not enough statistics below the mentioned lower fit limits, while above 100~fm event-by-event random fluctuations were present. A fit was deemed statistically acceptable if the confidence level, calculated from the $\chi^2$ of the fit and the number of degrees of freedom during the fit was above 0.1\%.

In all of the investigated cases, the measured $D(r_{\text{LCMS}})$ distribution had a power-law tail, which cannot be described by the Gaussian distribution. In order to achieve a precise description, the L\'evy distribution needs to be utilized. In case of CORE primordial pions without UrQMD and decays, the source shape was much closer to a Gaussian than in the other cases. After including the decay pions, but still not using UrQMD, the deviation from the Gaussian shape became larger, but the L\'evy fits were not statistically acceptable due to event-by-event fluctuations. In the cases where UrQMD was utilized, the L\'evy distribution provided a statistically acceptable description in most events for pions, kaons, and protons as well. Henceforward, these cases containing UrQMD were studied. Fits with a non-acceptable confidence level were discarded, but it is important to note that their fit parameter distributions were essentially identical to those of the fits with appropriate confidence levels.

\subsection{The distribution of the source parameters}

\begin{figure}[h!]
\centering
\includegraphics[trim=0 0.7cm 0 0.7cm, clip,scale=0.8]{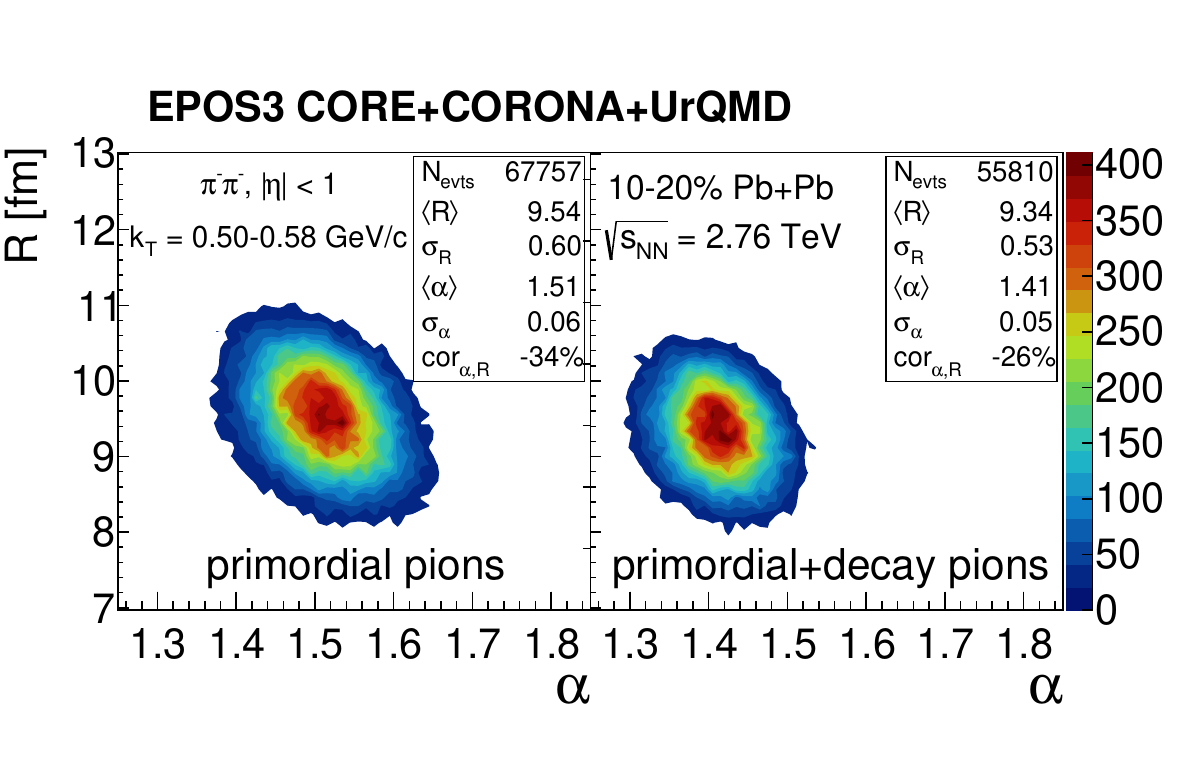}
\caption{The two-dimensional distribution of the L\'evy parameters $R$ and $\alpha$ for 10--20\% centrality, in the $\kT$ range of 0.50--0.58~GeV/$c$ for negatively charged primordial pions (left) and for negatively charged primordial and decay pions (right). The number of events ($N_{\text{evts}}$), the means, and the standard deviations of the distributions are shown in the legend. The scale on the right corresponds to the number of events in a single bin of the histogram.}
\label{f:Rvsalpha}
\end{figure}

The $R$ and $\alpha$ parameters of the L\'evy fits were collected in two-dimensional $R$ vs.\ $\alpha$ distributions for each centrality and $\kT$ class. An example distribution is shown in Fig.~\ref{f:Rvsalpha} for negatively charged pion pairs in the $\kT$ range of 0.50--0.58~GeV/$c$ and the centrality range of 10--20\%.

The mean L\'evy parameters $\left < R\right >$ and $\left < \alpha\right >$ for each centrality and $\kT$ class were calculated as the first moment of the $R$ vs.\ $\alpha$ distribution. The standard deviations $\sigma_R$ and $\sigma_{\alpha}$ were calculated as the second moment of the $R$ vs.\ $\alpha$ distribution. The $\left < R\right >$ and $\left < \alpha\right >$ values correspond to the expectation values of the parameters throughout many events, while $\sigma_R$ and $\sigma_{\alpha}$ indicate the fluctuation of the respective parameters among the events. The statistical uncertainties of $\left < R\right >$ and $\left < \alpha\right >$ are estimated by dividing $\sigma_R$ and $\sigma_{\alpha}$ with the square root of the number of events ($N_{\text{evts}}$) contributing to the individual $R$ vs.\ $\alpha$ distributions. As $N_{\text{evts}}$ has an order of magnitude of $10^4$, the statistical uncertainties are basically negligible.

\begin{figure}[h!]
\centering
\includegraphics[trim=0 0.7cm 0 0.7cm, clip,scale=0.8]{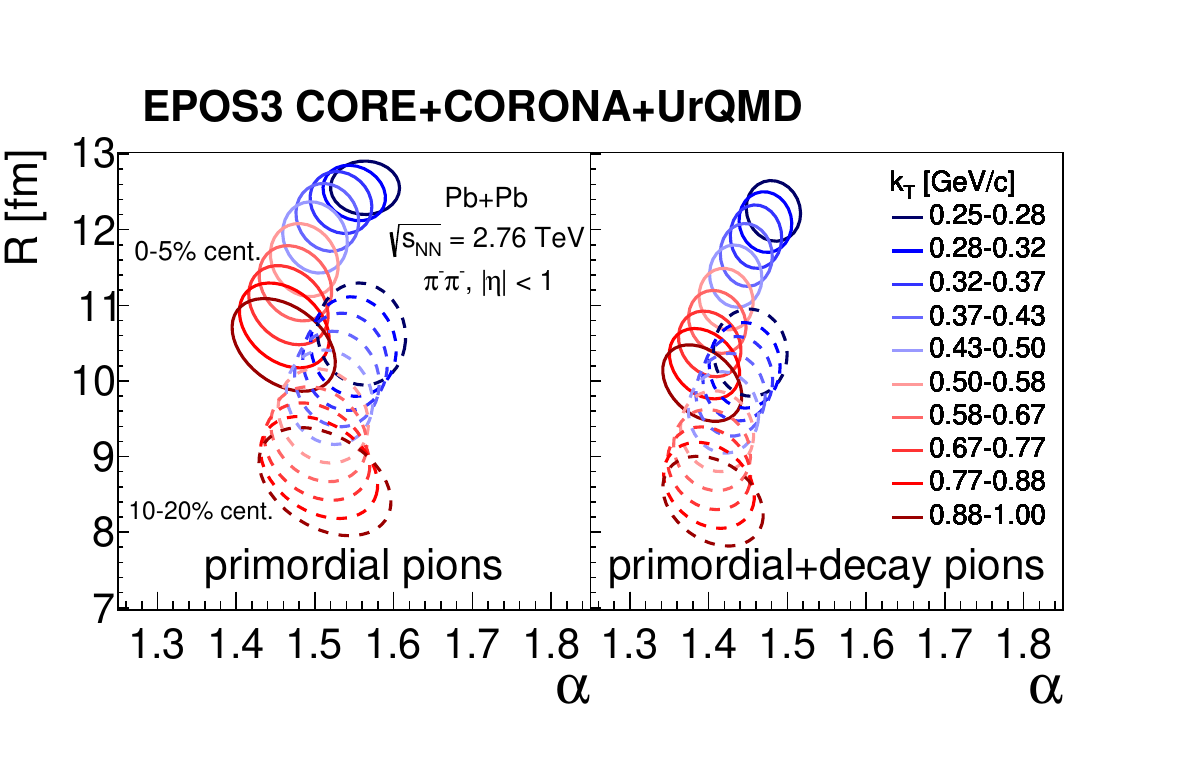}
\caption{The 1 $\sigma$ contours of the $R$ vs.\ $\alpha$ distributions in the different $\kT$ classes for two centralities for negatively charged primordial pions (left) and for negatively charged primordial and decay pions (right).}
\label{f:contours}
\end{figure}

To illustrate the centrality and $\kT$ dependence of the $R$ vs.\ $\alpha$ distributions, ellipses representing 1~$\sigma$ contours corresponding to the different centrality and $\kT$ classes are shown in Fig.~\ref{f:contours} for negatively charged pion pairs. The contour ellipses were calculated from $\sigma_R$, $\sigma_\alpha$ and the $\alpha-R$ correlation coefficient $\text{cor}_{\alpha,R}$. Out of the 4 used centrality classes only 2 are shown in order to maintain the clarity of the figure. The orientation of the contour ellipses shows that the $\alpha$ and $R$ parameters are anti-correlated.

The $R$ vs.\ $\alpha$ distributions for positively and negatively charged same-sign pion pairs were basically identical, and not much difference was observed in comparison with the opposite sign pion pairs either. Therefore, in each same centrality and $\kT$ class, the $R$ vs.\ $\alpha$ distributions corresponding to the three different charge combinations were combined. 

\section{Results and discussion}\label{s:results}

The mean L\'evy parameters $\left < R\right >$ and $\left < \alpha\right >$ were determined for pions, kaons and protons in different centrality and $\kT$ classes from an event-by-event analysis using the final stage of EPOS (CORE+CORONA+UrQMD). In case of pions, two separate cases were investigated, with and without the inclusion of decay products. For kaons and protons, only the case with the inclusion of the decay products was utilized. 

In this section, the centrality and the transverse mass ($\mT=\sqrt{\kT^2+m^2}$) dependence of $\left < R\right >$ and $\left < \alpha\right >$ is presented for pions, kaons and protons. The dependence on $\mT$, instead of $\kT$ is investigated to facilitate comparison with experimental results and with theory. The effect of the decay products is also analyzed in case of pions.

\subsection{The pion source parameters}

\begin{figure}[h!]
\centering
\includegraphics[trim=0 0 0 0, clip,scale=0.85]{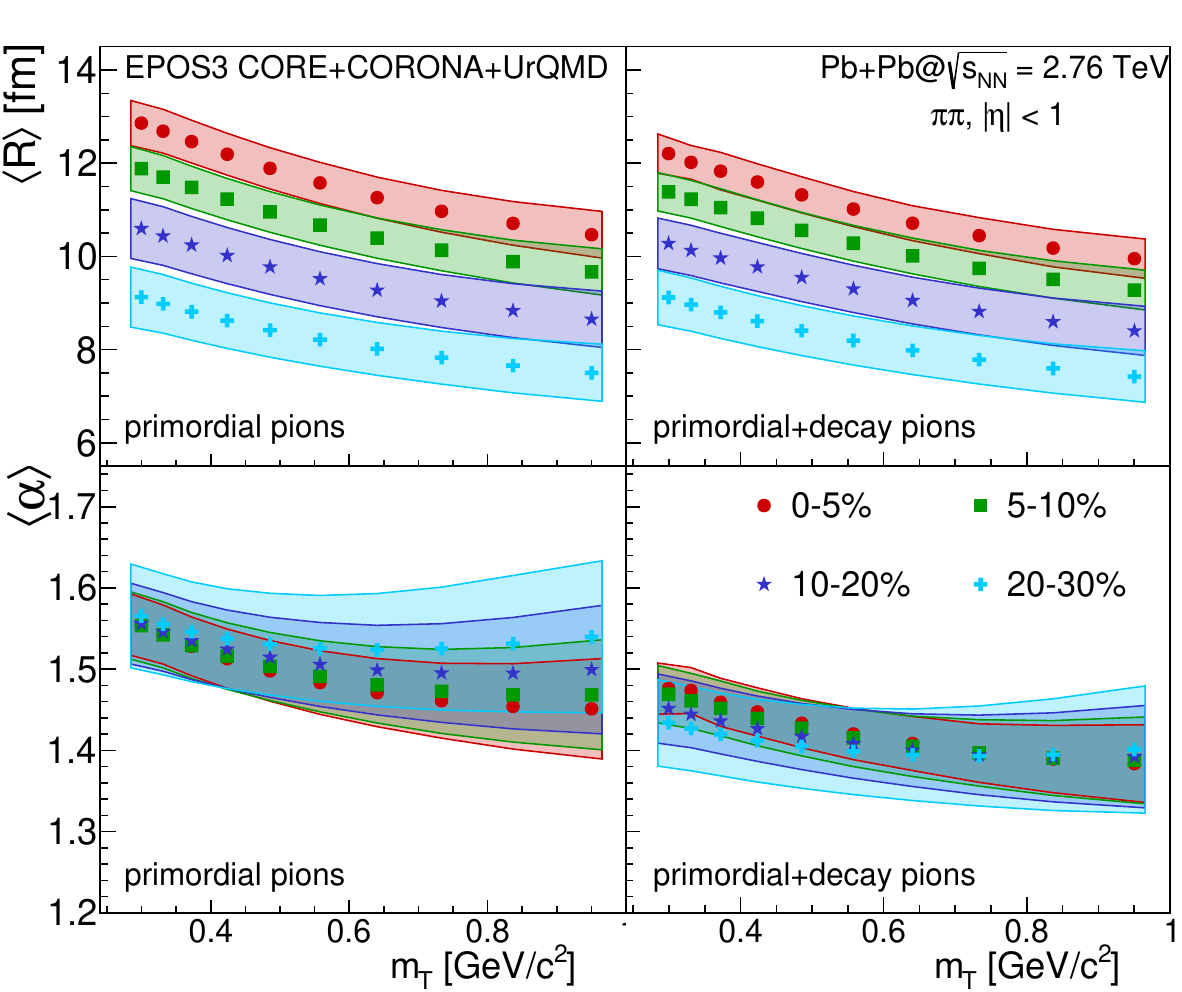}
\caption{The mean L\'evy scale parameter $\left < R\right >$ (upper row) and the mean L\'evy stability index $\left < \alpha\right >$ (lower row) vs.\ the transverse mass $\mT$ for the different centrality classes. The left column corresponds to the case of using primordial pions only, while the right column corresponds to the case of including both primordial and decay pions in the sample. The colored bands show the standard deviations. Statistical uncertainties are smaller than the marker size.}
\label{f:pionresults}
\end{figure}

The obtained $\left < R\right >$ and $\left < \alpha\right >$ parameters are shown in Fig.~\ref{f:pionresults} for pions. The $\left < R\right >$ parameter decreases with $\mT$ and as the collisions become more peripheral. The $\mT$ dependence is similar to the universally observed scaling of the Gaussian source-radii $R_{\text{G}}^{-2}\propto\mT$~\cite{Adler_2004,PHENIX:2009ilf,Adare_2018,CMS:2023xyd}, despite the source having a L\'evy distribution in this case. The centrality dependence shows the relation of $\left < R\right >$ to the initial size of the fireball. The effect of including decay pions is marginal for $\left < R\right >$, but for the most central classes, a small decrease of $\left < R\right >$ is observed. The centrality and the $\mT$ dependence of $\left < R\right >$ is further analyzed in Subsection~\ref{ss:pionkaonproton}, comparing the results for pions, kaons, and protons.

The $\left < \alpha\right >$ parameter has a less prominent $\mT$ and centrality dependence in case of pions as shown in Fig.~\ref{f:pionresults}. A small decrease with $\mT$ is observed both with and without the inclusion of the decay pions. In the former case, the $\mT$ dependence seems to change with centrality. The inclusion of the decay pions decreases $\left < \alpha\right >$ significantly, especially for the less central event class. This means that the decay products influence the shape of the source, resulting in a stronger tail, and thus a smaller $\left < \alpha\right >$.

It is important to compare the centrality and the $\mT$ dependence of the L\'evy source parameters $R$ and $\alpha$ to available experimental measurements, performed using unidentified hadrons by the CMS Collaboration in PbPb collisions at $\sqrt{s_{_{\text{NN}}}}=5.02$~TeV~\cite{CMS:2023xyd}. Furthermore, an event-by-event analysis with EPOS, similar to the current one, was also performed for pions using AuAu data at $\sqrt{s_{_{\text{NN}}}}=200$~GeV~\cite{Kincses:2022eqq}. The centrality and the $\mT$ dependence of $\left < R\right >$ and $\left < \alpha\right >$ observed in the present analysis is qualitatively similar to both the experimental results from CMS as well as the lower collision energy EPOS results. However, the particular value of the parameters in the same centrality and $\mT$ range is different. The $\left < R\right >$ values shown here are about 20--30\% larger than the lower energy EPOS values and are about twice as large as the experimentally measured values. The $\left < \alpha\right >$ values of 1.4--1.5 shown here in case of both primordial and decay pions are slightly smaller than the lower energy EPOS values of 1.5--1.6, and significantly lower than the experimental values of 1.7--1.9. There could be many reasons behind these differences, but their investigation is beyond the scope of the present paper. 

\subsection{Comparison of the pion, kaon and proton parameters} \label{ss:pionkaonproton}

\begin{figure}[h!]
\centering
\includegraphics[trim=0 0 0.3cm 0, clip,scale=0.7]{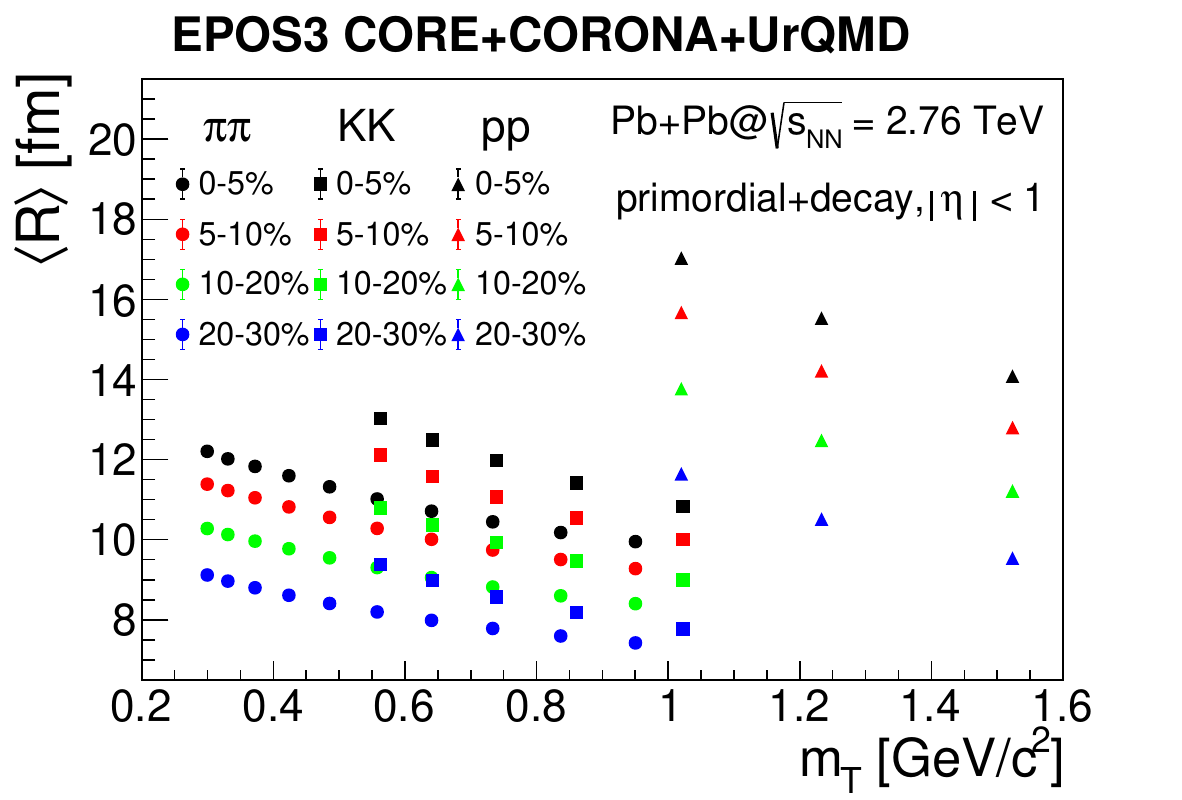}
\caption{The mean L\'evy scale parameter $\left < R\right >$ vs.\ the transverse mass $\mT$ for pions, kaons and protons in different centrality ranges using both primordial and decay particles. The error bars corresponding to the statistical uncertainties are smaller than the markers.}
\label{f:R}
\end{figure}

\begin{figure}[h!]
\centering
\includegraphics[trim=0 0 0.3cm 0, clip,scale=0.7]{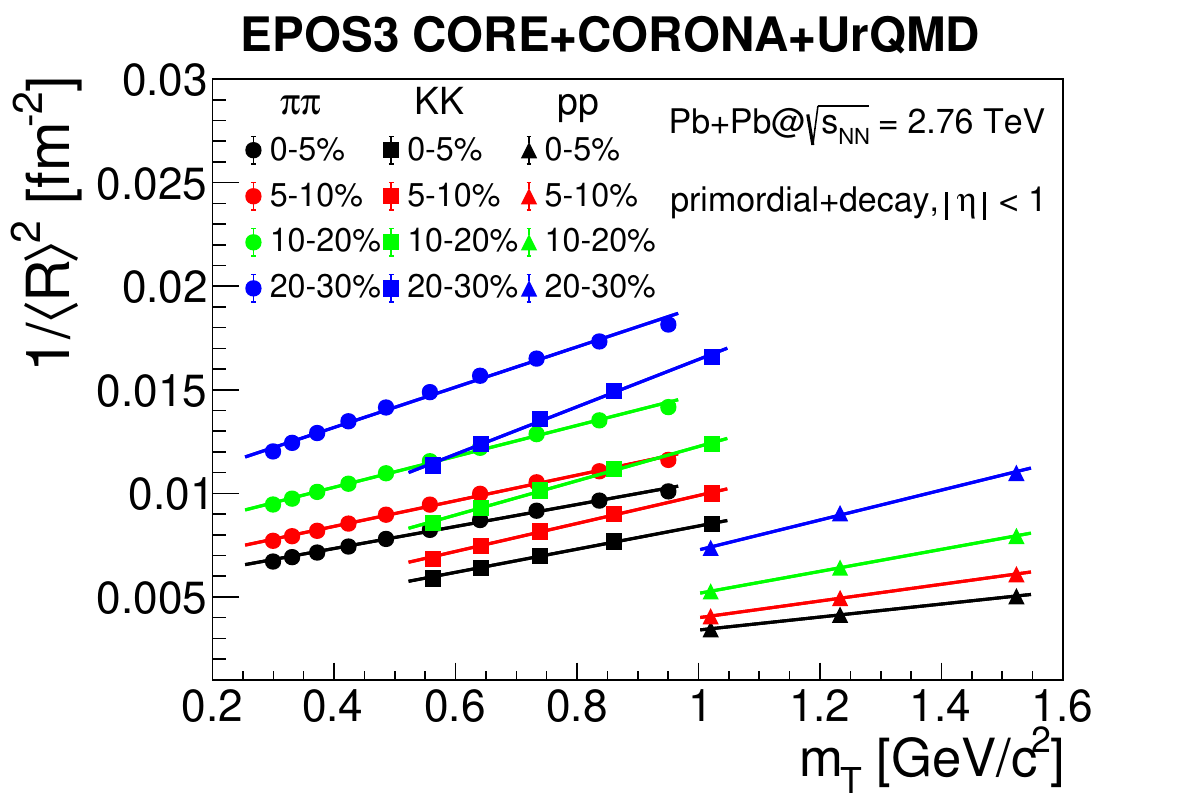}
\caption{The inverse square of the mean L\'evy scale parameter $1/\left < R\right >^2$ vs.\ the transverse mass $\mT$ for pions, kaons and protons in different centrality ranges using both primordial and decay particles. A line is fitted to the data for each centrality range and particle species. The error bars corresponding to the statistical uncertainties are smaller than the markers.}
\label{f:1_R2}
\end{figure}

The obtained $\left < R\right >$ parameters are shown in Fig.~\ref{f:R} for pions, kaons and protons. The decreasing trend with $\mT$, and as the collisions become more peripheral is also present for kaons and protons. Hydrodynamic models~\cite{makhlin1988hydrodynamics,Csorg__1996,Csanad:2008gt} predict that in case of a Gaussian source $R_{\text{G}}$ depends on the type of the particles only through $m_T$, with $R_{\text{G}}^{-2}\propto\mT$. This prediction is not fulfilled here, since $\left < R\right >$ is larger for kaons and even larger for protons in the same centrality and $\mT$ range. This behavior could be the result of the L\'evy source shape. However, for a given particle species, the $R^{-2}\propto\mT$ scaling is fulfilled, as shown in Fig.~\ref{f:1_R2}.

\begin{figure}[h!]
\centering
\includegraphics[trim=0 0 0.3cm 0, clip,scale=0.7]{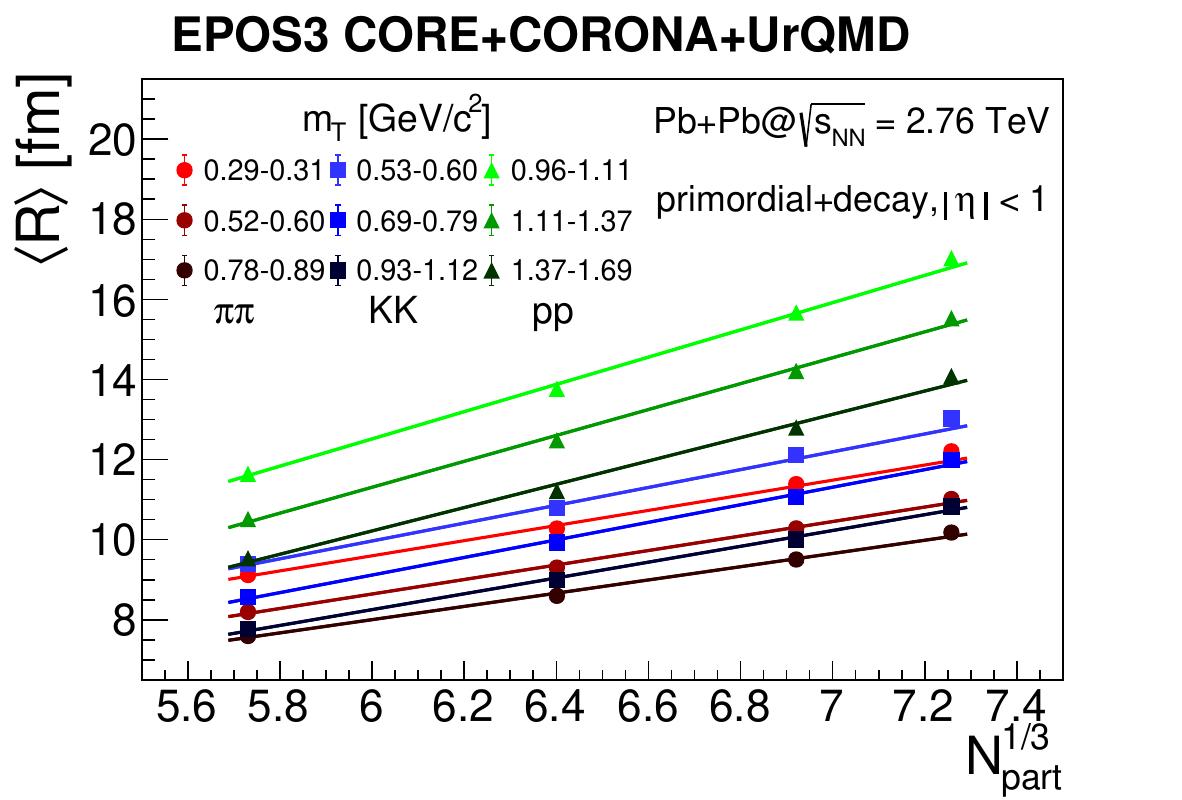}
\caption{The mean L\'evy scale parameter $\left < R\right >$ vs.\ the cubic root of the average number of nucleons participating in the collision $N_{\text{part}}^{1/3}$ for pions, kaons and protons in different transverse mass $\mT$ ranges using both primordial and decay particles. A line is fitted to the data for each $\mT$ range and particle species. The error bars corresponding to the statistical uncertainties are smaller than the markers.}
\label{f:RvsN}
\end{figure}

To further investigate the centrality dependence of $\left < R\right >$, it was plotted as a function of $N_{\text{part}}^{1/3}$, as shown in Fig.~\ref{f:RvsN}, for selected representative $\mT$ ranges, with $N_{\text{part}}$ being the average number of nucleons participating in the collision. The mapping from centrality to $N_{\text{part}}$ is given in Ref.~\cite{Loizides:2017ack}. A linear scaling of $\left < R\right >$ with $N_{\text{part}}^{1/3}$ is observed for all three investigated particle species, therefore $\left < R\right >$ is affine proportional to the initial size of the fireball.

\begin{figure}[h!]
\centering
\includegraphics[trim=0 0 0.3cm 0, clip,scale=0.7]{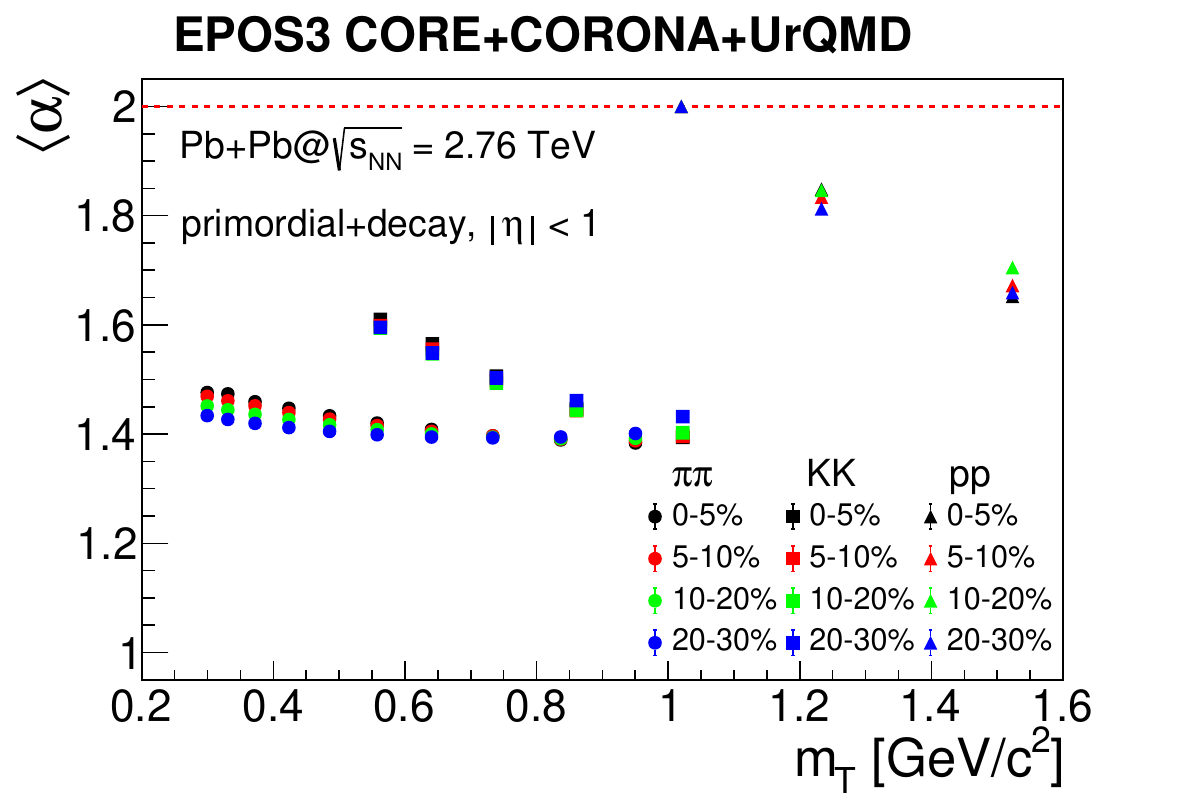}
\caption{The mean L\'evy stability index $\left < \alpha\right >$ vs.\ the transverse mass $\mT$ for pions, kaons and protons in different centrality ranges using both primordial and decay particles. The error bars corresponding to the statistical uncertainties are smaller than the markers.}
\label{f:alpha}
\end{figure}

The obtained $\left < \alpha\right >$ parameters are shown in Fig.~\ref{f:alpha} for pions, kaons and protons. The values are in the 1.4-1.5, 1.4-1.6 and 1.6-2.0 ranges for pions, kaons and protons, respectively. This means that the shape of the source is significantly different for the different species. For kaons and protons there is a clear decreasing trend with $\mT$, while this decrease is not so prominent for pions. The centrality dependence is quite small for all three particle species. If anomalous diffusion caused by hadronic rescattering would be the dominant reason behind the appearance of the L\'evy distribution, then $\left < \alpha\right >$ would be expected to be smaller for particles with a smaller total cross-section due to the larger mean free path~\cite{Csanad:2007fr}. This implies the relation $\alpha(p)>\alpha(\pi)>\alpha(K)$, as indicated in Ref.~\cite{Csanad:2007fr}. In the results obtained in the present analysis, the first inequality is fulfilled, but the second one is not. This implies that at least within the EPOS model, anomalous diffusion alone cannot be fully responsible for the L\'evy source shape. Note that experimental measurements for the L\'evy source parameters of kaons and protons are not yet published~\footnote{The authors are aware of the preliminary kaon results shown at the 15th Workshop on Particle Correlations and Femtoscopy, 2022; however, these are not final yet.} and there are no previous results from a phenomenological analysis either, except the source shapes shown in Ref.~\cite{Csanad:2007fr} noted above.

\begin{figure}[h!]
\centering
\includegraphics[trim=0 0 0.3cm 0, clip,scale=0.7]{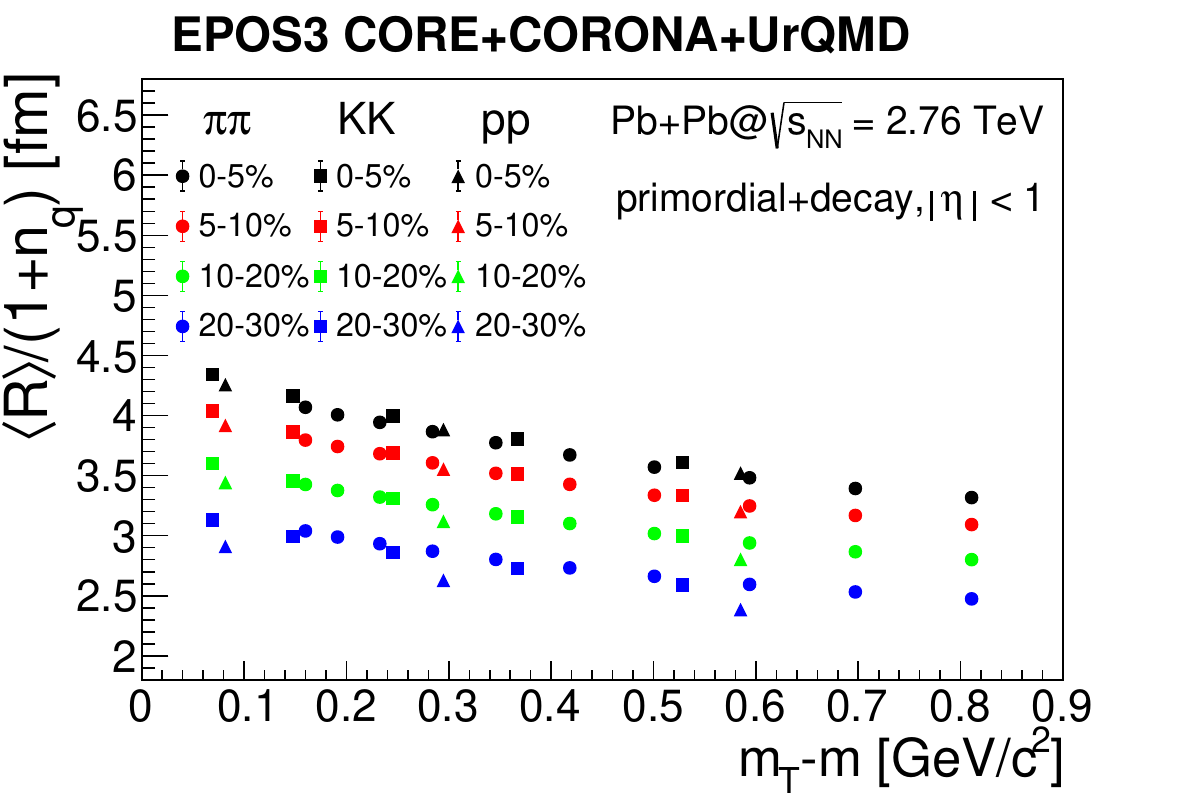}
\caption{The mean L\'evy scale parameter $\left < R\right >$ divided with one plus the number of valence quarks $n_{\text{q}}$ vs.\ the transverse kinetic energy $\mT-m$ for pions, kaons and protons in different centrality ranges using both primordial and decay particles. The error bars corresponding to the statistical uncertainties are smaller than the markers.}
\label{f:Rscaled}
\end{figure}

As discussed, $\left < R\right >$ does not follow a universal curve versus $\mT$ across the different particle species. However, it was found that by dividing $\left < R\right >$ with $1+n_{\text{q}}$, where $n_{\text{q}}$ is the number of valence quarks in the given particle species, and plotting the results as a function of $\mT-m$ (sometimes called ``transverse kinetic energy ''), a universal behavior independent of the particle species is obtained, as shown in Fig.~\ref{f:Rscaled}. This newly found scaling is not yet understood.

\section{Summary}\label{s:summary}

The investigation of the two-particle source function in lead-lead collisions simulated by the EPOS model at a center of mass energy per nucleon pair of \sqsntwo is presented. The two-particle source functions are directly reconstructed on an event-by-event basis for pions, kaons, and protons separately, using both primordial and decay particles in the final stage of EPOS. The L\'evy alpha-stable distribution (unlike the Gaussian distribution) is found to provide a good description of the two-particle source in the individual events for all three particle species. Hence it is clear that it is not the event averaging that leads to the appearance of L\'evy source shapes. The L\'evy scale parameter $R$ and the L\'evy stability index $\alpha$ are extracted as functions of pair average transverse momentum in each individual event. The means of the L\'evy parameters $\left < R\right >$ and $\left < \alpha\right >$ are determined in different collision centrality and pair average transverse mass ($\mT$) ranges. In case of pions, the effect of the decay products is also investigated by analyzing the two-particle source of only primordial pions and of primordial and decay pions as well.

In case of pions, a decreasing trend of $\left < R\right >$ with increasing $\mT$ is found, and similarly, $\left < R\right >$ decreases for more peripheral events. The centrality and the $\mT$ dependence of $\left < \alpha\right >$ is found to be less prominent. On the other hand, while $\left < R\right >$ is mainly unaffected by the inclusion of the decay pions, $\left < \alpha\right >$ is decreased, suggesting that the decay products influence the shape of the source and cause a heavier tail. The centrality and the $\mT$ dependent trends observed in the present analysis are similar to the trends observed in PbPb collisions at CMS and in lower energy EPOS simulation. However, the magnitude of $\left < R\right >$ is larger and the magnitude of $\left < \alpha\right >$ is smaller in the present analysis compared to both CMS data and lower energy EPOS simulation.

The hydrodynamic scaling of $\left < R\right >$ predicted in case of a Gaussian source is fulfilled for all three particle species separately. However, no universal $\mT$ scaling is observed across multiple particle species. The linear scaling of $\left < R\right >$ with the cubic root of the average number of nucleons participating in the collision is observed for all three particle species, confirming the relation of the $R$ parameter to initial geometry. The $\left < \alpha\right >$ parameter shows little $\mT$ dependence in case of pions, but a clear decreasing trend is observed in case of kaons and protons. The values of $\left < \alpha\right >$ are generally larger for kaons than for pions, and are even larger for protons. These trends only partially agree with the expectations in case of anomalous diffusion, suggesting that other reasons also contribute to the manifestation of the L\'evy source shape.

Most importantly, however, our results show that the particle emitting source is not Gaussian in individual EPOS events, but an event-by-event Lévy description is adequate. This calls for more detailed measurements of the Lévy stability exponent $\alpha$ at the LHC.

\section*{Acknowledgements}

The authors thank Johannes Jahan and Maria Stefaniak for useful discussions and providing simulated EPOS data. The authors also acknowledge support from the NKFIH grants TKP2021-NKTA-64, K-133046, K-128713, and K-138136. B. Kórodi was partly supported by the ÚNKP-22-2 New National Excellence Program of the Ministry for Culture and Innovation from the source of the National Research, Development and Innovation Fund.

\pagebreak
\printbibliography

\end{document}